\documentclass[twocolumn,notitlepage,tightenlines,showpacs,preprintnumbers,amsmath,amssymb]{revtex4}

\usepackage{hyperref}
\usepackage{graphicx,amsmath}
\usepackage{epsf}

\begin{document}
\title{Nonlinear Quantum Optics in Optomechanical Nanoscale Waveguides}
\author{Hashem~Zoubi}
\email{hashem.zoubi@itp.uni-hannover.de}
\author{Klemens~Hammerer}
\affiliation{Institute for Theoretical Physics, Institute for Gravitational
  Physics (Albert Einstein Institute), Leibniz University Hannover,
  Appelstrasse 2, 30167 Hannover, Germany}
\date{11 October 2016}

\begin{abstract}
We explore the possibility of achieving a significant nonlinear phase shift among
photons propagating in nanoscale waveguides exploiting interactions among photons that are
mediated by vibrational modes and induced through Stimulated Brillouin Scattering (SBS). We
introduce a configuration that allows slowing down the photons by several orders of
magnitude via SBS involving sound waves and two pump fields. We extract the conditions for maintaining vanishing amplitude gain or loss for slowly propagating photons while keeping the influence of thermal phonons to the minimum. The nonlinear phase among two counter-propagating photons can be used to realize a deterministic phase gate.
\end{abstract}


\maketitle

The non-interacting nature of photons makes them efficient as carriers for quantum information \cite{Walmsley2015} but non-efficient for information processing. Quantum nonlinear optics thrives to induce controlled interactions at the few photon level for fundamental physics and applications, e.g., for
photonic switches, memory devices and transistors
\cite{Chang2014,Reiserer2015,Firstenberg2016,Murray2016}. The ultimate challenge
is to achieve nonlinear phase shifts among two  optical photons realizing a quantum
logic gate for photonic quantum information processing
\cite{Imamoglu1997,OBrien2007,Kimble2008}. In the recent decades several directions have been
suggested for achieving effective photon-photon interactions. Among the first
experiments was Cavity Quantum Electrodynamics (CQED) using atoms as a
nonlinear medium \cite{Haroche2006,Reiserer2015},
which culminated in the recent demonstration of a deterministic quantum gate
\cite{Hacker2016} along the lines suggested in \cite{Poyatos1997}. Avoiding the use of resonators, strong nonlinearities have been achieved for
fields confined in waveguides, e.g., using tapered nanofibre strongly
coupled to an atomic chain \cite{Vetsch2010,Goban2012}. The restrictions on bandwidth imposed by the cavity spectrum motivated the
search for cavity free environments \cite{Hammerer2010}, for example, using Rydberg atoms in
a dense medium \cite{Hau2008,Gorshkov2011,Peyronel2012,Firstenberg2013} under the condition of Electromagnetic Induced
Transparency (EIT) \cite{Harris1990,Fleischhauer2005}, and later by exploiting the blockade phenomena \cite{Lukin2001,Pritchard2010}. The significant enhancement of photon-photon
interactions in the latter approach is mainly due to the achievement of slow
light using EIT which is subject to restrictions in bandwidth associated with the
transparency window \cite{Petrosyan2011}.

\begin{figure}
\includegraphics[width=\linewidth]{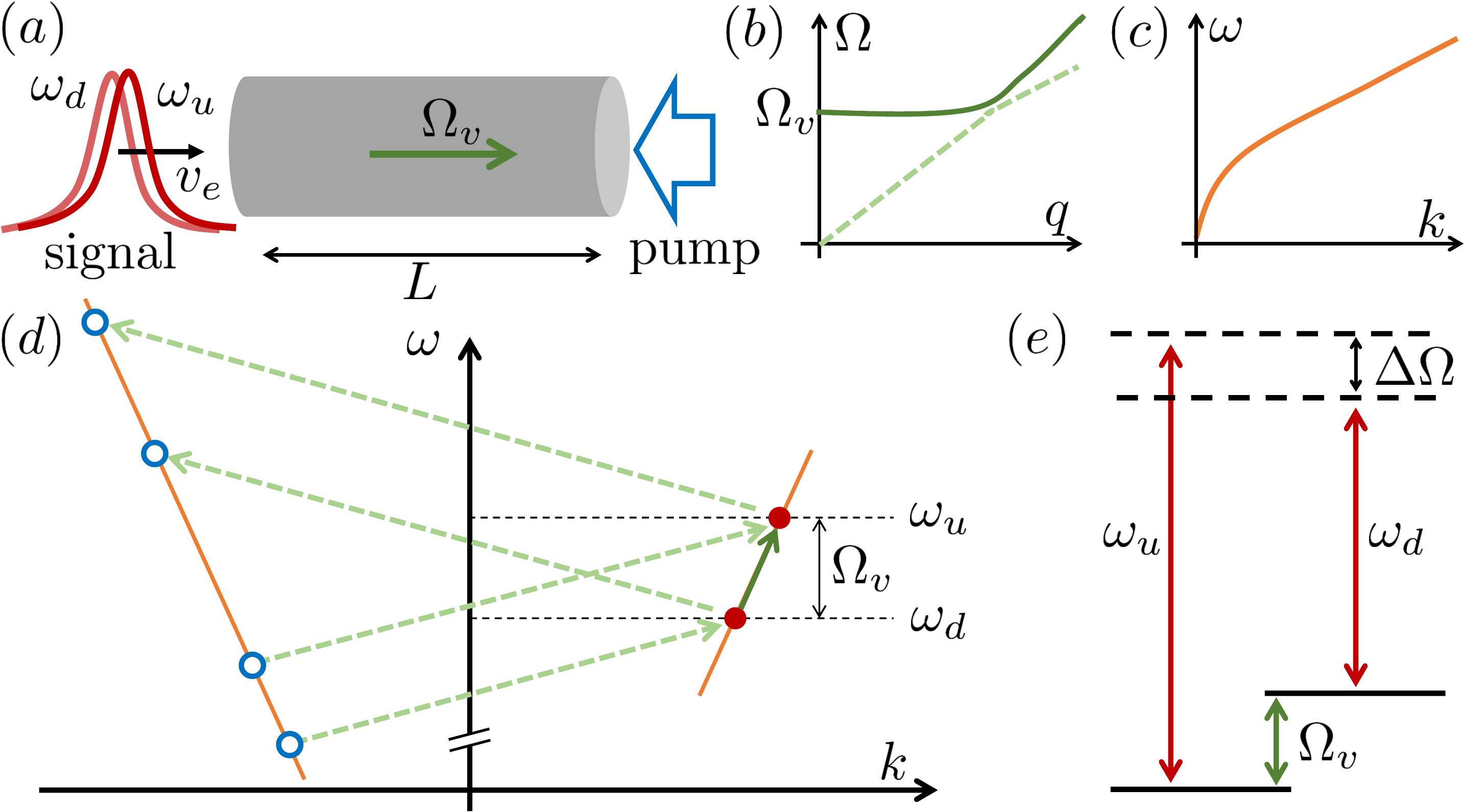}
\caption{(a) Schematic of the setup: Two signal fields at frequencies $\omega_{u(d)}$ propagate in a nanofibre of length $L$, and experience a cross-phase interaction mediated by SBS involving phonons of frequency $\Omega_v$. The effective group velocity $v_e$ is reduced due to EIT induced by counter-propagating pump fields.  (b) Schematic dispersion of the lowest two, dispersion-less (solid) and acoustic (dashed), phonon branches. (c) Schematic dispersion of the fundamental photon mode. (d) Zoom in on the photon dispersion: signal fields (solid circle) interact via dispersion-less phonons (solid arrow), four pump fields (empty circle) induce EIT at the signal frequencies via acoustic phonons (dashed arrows), cf. Fig.~\ref{Slow}. (e) Level scheme with detuning $\Delta\Omega$ between  signal fields and dispersion-less phonons.}
\label{NonPhase}
\end{figure}

In parallel, optical fibres \cite{Zhu2007,Thevenaz2008,Douglas2015,Goban2015} and photonic
crystals \cite{Russell2006,Baba2008,Eichenfield2009} have received significant interest,
as they can be easily integrated into all-optical on-chip platforms. In particular optical fibres can realize tunable delays of optical signals with the possibility of achieving fast and slow light in a comparatively wide bandwidth \cite{Okawachi2005,Song2005,Herraez2006}. The most efficient nonlinear process inside optical fibres is SBS, that is the scattering of optical photons by long lived
acoustic phonons commonly induced by electrostriction
\cite{Kim2015}. Recent progress in the fabrication of nanoscale waveguides in which the wavelength of light becomes larger than the waveguide
dimension achieved a breakthrough in SBS \cite{Pant2011,Shin2013,Eggleton2013}. In this regime the coupling of photons and phonons is significantly
enhanced due to radiation pressure dominating over electrostriction \cite{Rakish2012,VanLaer2015a} with significant implications for the field of Brillouin continuum optomechanics \cite{Hammerer2014,Rakich2016}.

In the present letter we introduce an efficient method for generating effective
interactions among photons induced through SBS
involving vibrational modes in nanoscale waveguides. Our scheme crucially relies on
achieving slow light by exploiting the significant scattering
of photons from acoustic phonons. We study the
correlations induced among slowly co- or counter-propagating photons, and show that a
significant nonlinear phase shifts can be accumulated along a cm scale
waveguide. We identify configurations where the slow group velocity of photons
can be exploited without net gain or loss in photon number which can be
achieved using two pump fields. We also consider the effect of thermal
fluctuations in the phonon modes and determine conditions for negligible
impact on the photon-photon interactions. Our treatment builds on the quantum mechanical Hamiltonian description of SBS in nanoscale waveguides recently developed in \cite{Sipe2016,Zoubi2016}. Quantum nonlinear optics and photon phase gates have been discussed previously in cavity optomechanics \cite{Rabl2011,Nunnenkamp2011,Stannigel2012,Ludwig2012,Wang2016} generally assuming a large single photon coupling (with the notable exception of \cite{Wang2016}). The results reported here relate to these previous schemes as the approach towards quantum nonlinear optics based on atomic ensembles relates to the one based on CQED.

We consider a cylindrical nanoscale waveguide of length $L$ on cm scale
with four pump fields propagating from right to left and two signal fields
containing few photons from left to right which are coupled through SBS to vibrational modes of the fibre, as represented in Figure
\ref{NonPhase}.a. The signal fields comprise wavenumbers centered around $k_u$
and $k_d$ of frequencies $\omega_u$ and $\omega_d$, respectively, as shown in Figures
\ref{NonPhase}.c and \ref{NonPhase}.d. The fields are described by slowly varying amplitude operators $\psi_\alpha(x)$ where $\alpha=u,d$. For an effectively one-dimensional photon field the real
space operator is expressed in terms of the momentum space one, $a_k$, by
$\psi_{\alpha}(x)=\frac{1}{\sqrt{L}}\sum_{k\in B_{\alpha}}a_k
e^{i(k-k_{\alpha})x}$. Here $B_{\alpha}$ denotes a suitable bandwidth of
photon wave numbers centered around a central wave numbers $k_{\alpha}$. The
definition of $\psi_{\alpha}(x)$ implies
$[\psi_{\alpha}(x),\psi_{\alpha}^\dagger(x')]=\delta(x-x')$
where the $\delta$-function is understood to be of width $\sim
B^{-1}_{\alpha}$. Moreover, we consider (effectively) dispersion-less vibrational modes of frequency $\Omega_v$ and
wavenumber $q_v$ which are represented by a slowly varying
phonon field operator $Q(x)$, as appeared in Fig.~\ref{NonPhase}.b. The two photonic signal modes are
detuned from the vibration by $\Delta\Omega=\omega_u-\omega_d-\Omega_v$
with difference in wavenumbers of $\Delta q=k_u-k_d-q_v$, cf. Fig.~\ref{NonPhase}.e. The two signal fields are assumed to propagate at a slow group velocity $v_e$ which can be achieved by a proper choice of pump fields  exploiting SBS involving acoustic phonons, as will be explained in detail further below. The Hamiltonian for the two slow signal fields and the vibrational modes reads \cite{Zoubi2016} $(\hbar=1)$
\begin{align}
H&=H_0-i v_e\sum_{\alpha}\int
dx\ \psi_{\alpha}^{\dagger}(x)\frac{\partial\psi_{\alpha}(x)}{\partial
  x} \\
&\quad+\sqrt{L}\int dx\ \left(f_v\ Q^{\dagger}(x)\psi_d^{\dagger}(x)\psi_u(x)\ e^{i\Delta q x}+h.c.\right)\nonumber,
\end{align}
where $H_0=\sum_{\alpha}\int
dx\omega_\alpha\psi_{\alpha}^{\dagger}(x)\psi_{\alpha}(x)+\Omega_v\int
dxQ^{\dagger}(x)Q(x)$. The frequency $f_v$ describes the strength of SBS
among the two photonic signal fields and the vibrational fields. In the
local field approximation it is independent of the wavenumber. The corresponding equations of motion for the photon operators in an interaction picture with respect to $H_0$ are
\begin{align}
\Big(\tfrac{\partial}{\partial t}+v_e&\tfrac{\partial}{\partial
  x}\Big)\psi_u(x,t)=\nonumber \\
& -if_v\sqrt{L}\ Q(x,t)\psi_d(x,t)\ e^{i(\Delta\Omega t-\Delta qx)}, \nonumber \\
\Big(\tfrac{\partial}{\partial t}+v_e&\tfrac{\partial}{\partial  x}\Big)\psi_d(x,t)=\nonumber \\
& -if_v\sqrt{L}\   Q^{\dagger}(x,t)\psi_u(x,t)\ e^{-i(\Delta\Omega t-\Delta qx)}.\label{EoM:amplitudes}
\end{align}
The phonon operator evolves as
\begin{align}
\Big(\tfrac{\partial}{\partial
  t}+&\tfrac{\Gamma_v}{2}\Big)Q(x,t)=\\
&-if_v\sqrt{L}\ \psi_d^{\dagger}(x,t)\psi_u(x,t)\ e^{-i(\Delta\Omega
  t-\Delta qx)}-{\cal F}(x,t),\nonumber
\end{align}
where $\Gamma_{v}$ is the vibrational mode
damping rate, and ${\cal F}(x,t)$ is the Langevin
noise operator \cite{Boyd1990} fulfilling $[{\cal F}^\dagger(x,t),{\cal
    F}(x',t')]=\Gamma_{v}\delta(x-x')\delta(t-t')$ and $\langle
{\cal F}(x,t){\cal
  F}^\dagger(x',t')\rangle=\Gamma_{v}(\bar{n}_{v}+1)\delta(x-x')\delta(t-t')$,
where $\bar{n}_{v}$ is the average number of thermal phonons. We assumed that
photon loss is negligible on the time scale $L/v_e$ of propagation of photons
through the fibre. Dominant photon loss is to be expected from in- and out-coupling of photons from the nanofibre.

We will show now that the two signal fields experience a significant cross-phase shift mediated through their off-resonant interaction with the vibrational field. For sufficiently large detuning $\Delta\Omega>f_v$ the phonon field can be adiabatically eliminated from the equations of motion \eqref{EoM:amplitudes} giving rise to a closed set of equations for the photon fields which can be integrated thanks to an (approximate) conservation of the number of photons in each mode.
In order to demonstrate this we define the photon number density $\hat{N}_{\alpha}(x,t)=\psi_{\alpha}^{\dagger}(x,t)\psi_{\alpha}(x,t)$ for mode $\alpha=u,d$ and the total photon density $\hat{N}=\hat{N}_u+\hat{N}_d$. Direct calculation using the change of variables $\xi=x-v_{e}t$ and
$\eta=v_{e}t$, after adiabatic elimination of the phonons, yields $\frac{\partial}{\partial
  \eta}\hat{N}(\xi,\eta)=0$. For the time being we
drop the Langevin term, and consider its influence in much details
later. Thus the total photon density is conserved during propagation through
the fibre, $\hat{N}^\mathrm{out}(\xi)=\hat{N}^\mathrm{in}(\xi)$, where we use
the definition of input and output operators
$\hat{\mathcal{O}}^\mathrm{in[out]}(\xi)=\hat{\mathcal{O}}(\xi,0[L])$ for any
observable $\hat{\mathcal{O}}(\xi,\eta)$. Moreover, one finds that the photon
number densities $\hat{N}_{\alpha}(\xi,\eta)$ obey the Riccati equations \cite{SupMat}
\begin{align}
\frac{\partial}{\partial
  \eta}\hat{N}_u(\xi,\eta)&=-V\hat{N}(\xi)\ \hat{N}_u(\xi,\eta)+V\ \hat{N}_u^2(\xi,\eta),
\nonumber \\
\frac{\partial}{\partial \eta}\hat{N}_d(\xi,\eta)&=V\hat{N}(\xi)\ \hat{N}_d(\xi,\eta)-V\ \hat{N}_d^2(\xi,\eta).
\end{align}
Here $V=\vartheta\frac{\Gamma_v/(\Delta\Omega)}{1+\Gamma_v^2/(4\Delta\Omega^2)}$, and $\vartheta=\frac{f_v^2L}{v_e\Delta\Omega}$ will turn out to be the nonlinear phase shift among the modes $u$ and $d$, see below. The input-output relations resulting from these equations are $\hat{N}_u^{\mathrm{out}}(\xi)=\hat{N}_u^{\mathrm{in}}(\xi)\hat{N}^\mathrm{in}(\xi)\big[\hat{N}_u^{\mathrm{in}}(\xi)+e^{VL\hat{N}_\mathrm{in}(\xi)}\hat{N}_d^{\mathrm{in}}(\xi)\big]^{-1}$ and $\hat{N}_d^{\mathrm{out}}(\xi)=\hat{N}_d^{\mathrm{in}}(\xi)\hat{N}^\mathrm{in}(\xi)\big[\hat{N}_d^{\mathrm{in}}(\xi)+e^{-VL\hat{N}_\mathrm{in}(\xi)}\hat{N}_u^{\mathrm{in}}(\xi)\big]^{-1}$ where we used that input number density operators commute. For input states in the signal modes which fulfill $VL\langle\hat{N}_\mathrm{in}(\xi)\rangle\ll1$ the photon number in each mode is conserved, $\hat{N}^\mathrm{out}_\alpha(\xi)=\hat{N}^\mathrm{in}_\alpha(\xi)$, as we will assume in the following. It is interesting to note that in the opposite case the nonlinear interaction of photons acts as an incoherent adder in mode $d$, that is $\hat{N}_d^{\mathrm{out}}(\xi)=\hat{N}^{\mathrm{in}}(\xi)$, while $\hat{N}_u^{\mathrm{out}}(\xi)=0$.

In the limit where both $\hat{N}_u$ and $\hat{N}_d$ are conserved during their
propagation in the waveguide the input-output relations for the photon field operators are \cite{SupMat}
\begin{subequations}\label{InOutAmpl}
\begin{align}
\psi^\mathrm{out}_u(\xi)&=\psi^\mathrm{in}_u(\xi)e^{-i\vartheta \hat{N}^\mathrm{in}_d(\xi)L}\\
&+\frac{i}{v_e}\int_0^{L}d\eta'\ U(\xi,\eta')\ \psi_d(\xi,\eta')e^{-i\vartheta \hat{N}^\mathrm{in}_d(\xi)\ (L-\eta')},
\nonumber \\
\psi^\mathrm{out}_d(\xi)&=\psi^\mathrm{in}_d(\xi)e^{-i\vartheta \hat{N}^\mathrm{in}_u(\xi)L} \\
&+\frac{i}{v_e}\int_0^{L}d\eta'\ U^\dagger(\xi,\eta')\ \psi_u(\xi,\eta')e^{-i\vartheta \hat{N}^\mathrm{in}_u(\xi)\ (L-\eta')},\nonumber
\end{align}
\end{subequations}
with $U(x,t)=f_v\sqrt{L}\ e^{i(\Delta\Omega t-\Delta
  qx)}\int_{0}^t dt'\ {\cal F}(x,t')e^{-\frac{\Gamma_v}{2}(t-t')}$. In the first line of both Equations~\eqref{InOutAmpl} the nonlinear
cross-phase shift $\vartheta$ appears in the exponent. The second lines
describe the contributions due to thermal fluctuations of the phonon modes
which generate an incoherent mixing of photon field amplitudes in modes $u$
and $d$. Using the properties of the Langevin force operators the average
number of photons at the waveguide output are given by
$N^\mathrm{out}_u=N_u^\mathrm{in}+N_u^\mathrm{fluct}$, and $N^\mathrm{out}_d=N_d^\mathrm{in}+N_d^\mathrm{fluct}$, where
$N_{\alpha}=\langle\hat{N}_{\alpha}\rangle$. The average number of
incoherently added photons is \cite{SupMat} $N^\mathrm{fluct}_u\approx W\bar{n}_vN^\mathrm{in}_d$ and
$N^\mathrm{fluct}_d\approx W(1+\bar{n}_v)N^\mathrm{in}_u$ where $W=\frac{L^3 \Gamma_vf_v^2}{v_e^3}$. For $W(1+\bar{n}_v)\ll 1$ incoherently added photons make a small relative contribution.

As an example we consider a cylindrical nanofibre where $f_v=3.3\times10^6$~Hz can be achieved for $L=1$~cm and a diameter $d=500$~nm in silicon, as we have shown in \cite{Zoubi2016}. At the same time
one finds $\Omega_v=2\pi\times 10$~GHz for longitudinal modes such that $\bar{n}_v\approx
0.1$ at $T=200$~mK. For a detuning of $\Delta\Omega=5\times 10^6~\mathrm{Hz}>f_v$ one obtains a
significant nonlinear phase shift of $\vartheta\approx 1$ for an effective group velocity  of $v_e\approx 2.2\times 10^4$~m/s which is reachable in this system as discussed below. In order to guarantee a small number of incoherent excitations $W\approx 0.1$ one has to require a mechanical quality factor $Q_v=6\times 10^5$ (that is, $\Gamma_v=10^5$~Hz). At the same time, this implies $V\simeq 0.02$ such that the number of photons in each mode is conserved as long as the input photon flux fulfills $v_e\langle\hat{N}_\mathrm{in}(\xi)\rangle \ll \frac{v_e}{VL}\simeq 10^8$\,sec$^{-1}$. The acceptable bandwidth of photons in the two modes $u$ and $d$ has to be small on the scale of the detuning but may be still on the order of $500$~kHz. We emphasize that the nonlinear phase resulting from these parameters is of the same order as the one achieved using cold atoms exploiting the Rydberg blockade phenomenon.

The nonlinear phase shifts appearing in Eqs.~\eqref{InOutAmpl} can be viewed formally as arising from a cross-phase interaction Hamiltonian $H_\mathrm{eff}=gL\int dx\ \psi_u^{\dagger}(x)\psi_d^{\dagger}(x)\psi_u(x)\psi_d(x)$ among the two photons where
$g=\frac{f_v^2}{\Delta\Omega}$. This interaction gives rise to a wide range of
quantum nonlinear optics on the level of single photons and many-body physics
of photons \cite{Chang2014}. For two co-propagating single-photon pulses the nonlinear phase shift comes along with changes and correlations in the spatio-temporal profile of the pulses \cite{Shapiro2006} limiting the applicability of the nonlinear
phase shift for the implementation in two-qubit quantum logic gate \cite{Fleischhauer2005,Gea-Banacloche2010}. This
effect can be suppressed by using counter-propagating pulses that still experience an identical nonlinear phase \cite{Gorshkov2011}. For counter-propagating modes $u$ and $d$ the treatment is essentially equivalent to the one given above and results in the same effective Hamiltonian $H_\mathrm{eff}$. Solving the Schr\"{o}dinger equation  for an initial state of two incoming counter-propagating photons in the waveguide $|\tilde{\phi}\rangle_\mathrm{in}=\int
dx_1dx_2\ \phi(x_1,x_2,t)\ \psi^{\dagger}_u(x_1)\psi^{\dagger}_d(x_2)|\mathrm{vac}\rangle$,
where $\phi(x_1,x_2,t)$ is a given two-photon wave function, it is straight forward to derive the scattering relation $|\tilde{\phi}\rangle_\mathrm{out}=e^{i\vartheta}|\tilde{\phi}\rangle_\mathrm{in}$. A unique application of such nonlinear phase shift among single photons is in all-optical deterministic quantum logic.

\begin{figure}
\includegraphics[width=\linewidth]{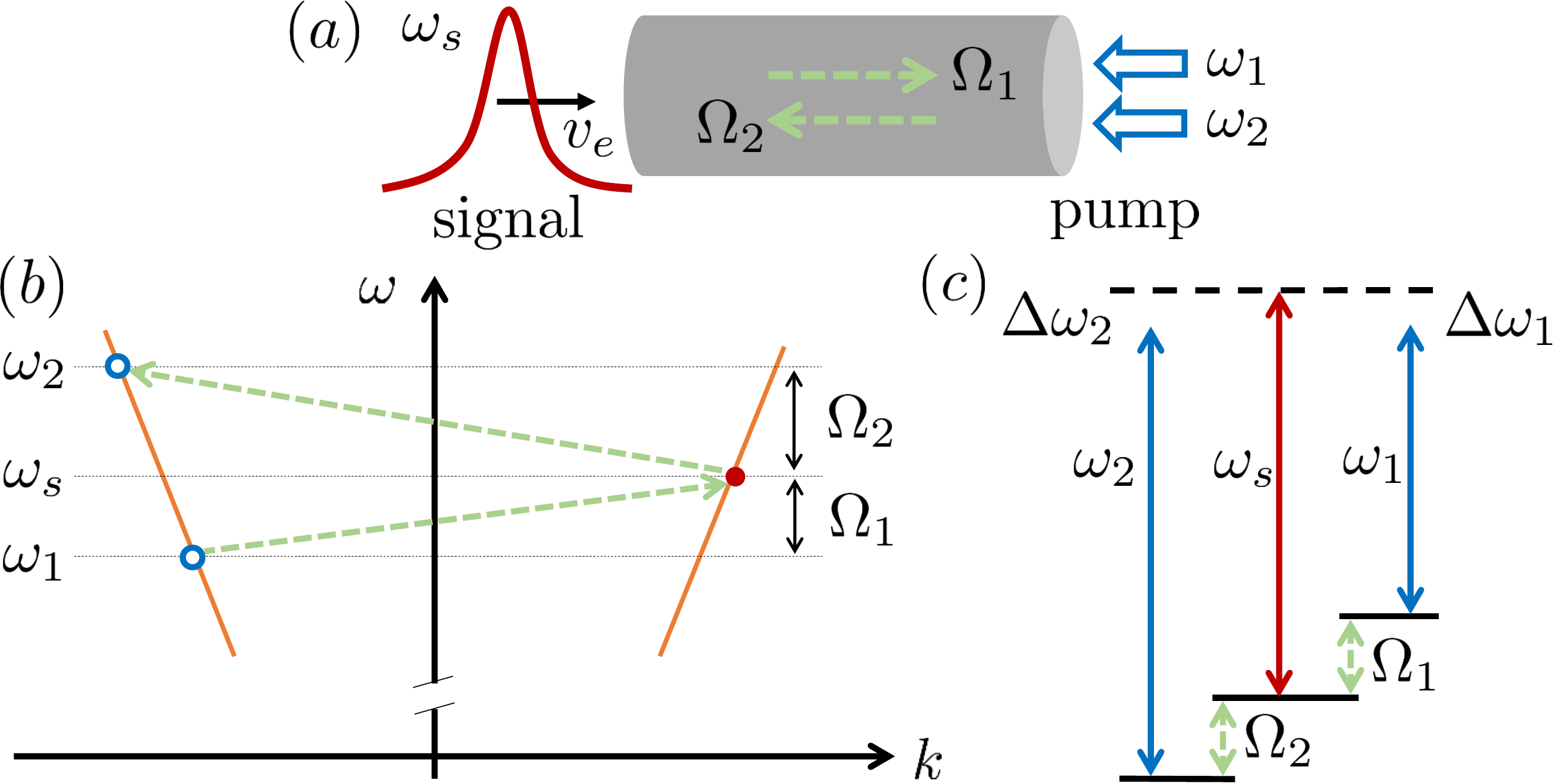}
\caption{Scheme for achieving gain- and lossless slow light: (a) A signal field at frequency $\omega_s$ is dressed by two pump fields $\omega_{1(2)}$ interacting through acoustic phonons at frequencies $\Omega_{1(2)}$.  (b) Photon dispersion with signal field (solid circle), pump fields (empty circles) and resonant acoustic phonon modes (dashed arrows). (c) Schematic level scheme with detunings $\Delta\omega_1=\omega_s-\omega_1-\Omega_1$ and $\Delta\omega_2=\omega_2-\omega_s-\Omega_2$
  among the fields.}
\label{Slow}
\end{figure}

In order to observe sizable nonlinear phase shifts it is crucial to achieve a
small effective group velocity $v_e$. Slow (or fast) light based on SBS in
waveguides has been demonstrated in several experiments
\cite{Herraez2005,Chin2006,Zhu2006}, and similar results have been achieved in
cavity optomechanics \cite{Weis2010,Safavi-Naeini2011,Kim2015}. The effect can be understood in analogy
to EIT in atomic media where acoustic phonons play the role of internal atomic
states. Slowing (advancing) of light based on SBS in general is linked to a
net Brillouin gain (loss) in the signal field \cite{Thevenaz2008} which, in a
quantum mechanical treatment, will be connected necessarily to additional
noise affecting the signal field. Therefore, in order to exploit SBS induced
slowing of  light for quantum nonlinear optics it is crucial to suppress
Brillouin gain or loss while maintaining a slow group velocity. We will show
now how this can be achieved using two pump fields counter-propagating to the
signal field. We consider a signal field of frequency $\omega_s$ and wave number $k_s$
propagating to the right with group velocity $v_g$ that is described by the operator $\psi(x,t)$. Two additional strong (classical) fields of frequencies
$\omega_1$ and $\omega_2$ with wavenumbers $k_1$ and $k_2$ are propagating to
the left with the same group velocity $v_g$, as shown in Figures \ref{Slow}.a
and \ref{Slow}.b. The signal is detuned from
the sum of field $(1)$ and a phonon of frequency $\Omega_1$
 by the detuning
$\Delta\omega_1=\omega_s-\omega_1-\Omega_1$. On the other hand, field $(2)$ is
detuned from the sum of the signal and a phonon of frequency
$\Omega_2$ by the detuning
$\Delta\omega_2=\omega_2-\omega_s-\Omega_2$, cf. Fig.~\ref{Slow}.c. The two acoustic phonons are described
by the operators $Q_1(x,t)$ and $Q_2(x,t)$, with sound velocity $v_a$ and
wavenumbers $q_1$ and $q_2$, respectively. The strong fields $(1)$ and $(2)$ are
taken to be classical amplitudes ${\cal E}_1(x,t)$ and ${\cal
  E}_2(x,t)$, which are defined by ${\cal E}_{\alpha}(x,t)=\sqrt{L}\langle\psi_p^{\alpha}\rangle$. The
configuration of fields is shown in Fig.~\ref{Slow}.b, and for the case
of two signals in Fig.~\ref{NonPhase}.d. The system is described by
\begin{align}
H&=-i v_g\int
dx\ \psi^{\dagger}(x)\frac{\partial\psi(x)}{\partial
  x} \nonumber \\
&\quad-i v_a\int
dx\left\{  Q_1^{\dagger}(x)\frac{\partial Q_1(x)}{\partial
  x}-Q_2^{\dagger}(x)\frac{\partial Q_2(x)}{\partial
  x}\right\} \nonumber \\
&\quad+\int dx\ \left(f^{a}_{1}{\cal E}_{1}^{\ast}(x)\ Q_{1}^{\dagger}(x)\psi(x)+h.c.\right) \nonumber \\
&\quad+\int dx\ \left(f^{a}_{2}{\cal E}_{2}^{\ast}(x)\ Q_{2}(x)\psi(x)+h.c.\right).
\end{align}
As before we assume the photon and phonon dispersions to be linear with group velocities $v_g$ and $v_a$, respectively. The photon-phonon coupling parameters, $f^a_1$ and $f^a_2$ are taken in the local
field approximation. The acoustic phonons have a damping rate of $\Gamma_a$, and the photons have a negligible damping. Thermal fluctuations of phonons are included by adding Langevein noise operators, ${\cal F}_i(x,t)$ \cite{Boyd1990}. The equation of motion for signal photons reads
\begin{align}
\Big(\frac{\partial}{\partial t}+v_g\frac{\partial}{\partial
  x}\Big)&\psi(x,t)=\\
&=-i f^a_1{\cal
  E}_{1}(x,t)\ Q_1(x,t)\ e^{i(\Delta\omega_1t-\Delta k_1x)} \nonumber \\
&\quad-i f^a_2{\cal
  E}_{2}(x,t)\ Q_2^{\dagger}(x,t)\ e^{-i(\Delta\omega_2t+\Delta
    k_2x)},\nonumber
\end{align}
and the ones for the phonon modes are
\begin{align}
\Big(\frac{\partial}{\partial t}&+v_a\frac{\partial}{\partial
  x}+\frac{\Gamma_a}{2}\Big)Q_{1}(x,t) \\
  &=-i f_1^{a}{\cal E}_{1}^{\ast}(x,t)\ \psi(x,t)\ e^{-i(\Delta\omega_1t-\Delta k_1x)}-{\cal F}_{1}(x,t),
\nonumber \\
\Big(\frac{\partial}{\partial t}&-v_a\frac{\partial}{\partial
  x}+\frac{\Gamma_a}{2}\Big)Q_{2}(x,t)\nonumber \\
  &=-i f^a_2{\cal E}_{2}(x,t)\ \psi^{\dagger}(x,t)\ e^{-i(\Delta\omega_2t+\Delta k_2x)}-{\cal F}_{2}(x,t),\nonumber
\end{align}
where $\Delta k_1=k_s+k_1-q_1$ and $\Delta k_2=k_2+k_s-q_2$.

Elimination of the acoustic phonons leads to the formal solution of the photon operator \cite{SupMat}
\begin{align}\label{Signal}
\psi&(x,t)=e^{-({G}+i\kappa)x}\psi_\mathrm{in}(x-v_gt)+i\frac{f_a}{v_g}\int_0^{x}dx'e^{({G}+i\kappa)(x'-x)}\nonumber \\
&\times\left\{{\cal E}_{1}e^{-i\Delta k_1x'}W_1(x',t)+{\cal E}_{2}e^{-i\Delta k_2x'}W_2^{\dagger}(x',t)\right\},
\end{align}
where $\psi_\mathrm{in}(x-v_gt)$ is the incident signal operator. We defined the gain coefficient ${G}=\frac{f_a^2\Gamma_a}{2v_g}\left\{\frac{|{\cal
  E}_{1}|^2}{\frac{\Gamma_a^2}{4}+\Delta\omega_1^2}-\frac{|{\cal
  E}_{2}|^2}{\frac{\Gamma_a^2}{4}+\Delta\omega_2^2}\right\}$, the shift in wave number $\kappa=\frac{f_a^2}{v_g}\left\{\frac{|{\cal
  E}_{1}|^2\Delta\omega_1}{\frac{\Gamma_a^2}{4}+\Delta\omega_1^2}+\frac{|{\cal
  E}_{2}|^2\Delta\omega_2}{\frac{\Gamma_a^2}{4}+\Delta\omega_2^2}\right\}$, and the noise operators
$W_i(x,t)=e^{i\Delta\omega_it}\int_{0}^t
dt'\ {\cal F}_{i}(x,t')e^{-\frac{\Gamma_a}{2}(t-t')}$. For simplicity we assumed $f^a_1=f^a_2\equiv f_a$ and constant pumps.

At this point we estimate the contribution of the thermal fluctuations. We calculate the average number of signal photons using the properties of Langevin noise operators given above \cite{Boyd1990}. We are interest in the limit of negligible gain, that is $GL\ll1$. Later we extract the condition for achieving this limit. The thermal
photons appear due to the scattering of the upper pump photons into the signal
photons which is induced by thermal phonons. In the
limit considered here we get $\hat{N}_\mathrm{out}=\hat{N}_\mathrm{in}+\hat{N}_\mathrm{fluct}$, where the density of  incident photons is
$N_\mathrm{in}=\langle\psi_\mathrm{in}^{\dagger}(L-v_gt)\psi_\mathrm{in}(L-v_gt)\rangle$,
and the average
number of incoherently added photons is \cite{SupMat} $\langle\hat{N}_\mathrm{fluct}\rangle \approx\frac{f_a^2\Gamma_a L^2}{v_g^3}\left\{|{\cal
  E}_{1}|^2\bar{n}_a^{(1)}+|{\cal E}_{2}|^2(\bar{n}_a^{(2)}+1)\right\}$, where $\bar{n}_a^{(i)}$ is the average number of thermal phonons in the reservoir at
frequency $\Omega_i$. For phonons of frequency $\Omega_a=2\pi\times 15$~GHz, at temperature $T=200$~mK, the average
number of thermal phonons is $\bar{n}_a\approx 0.03$. Using the numbers
$f_a=2.6\times 10^5$~Hz, $L=1$~cm, $|{\cal
    E}_1|^2=4\times10^8$, and $|{\cal
    E}_2|^2=10^8$, which
are equivalent to about $10$~mW, we get a number density of incoherent photons of $\langle\hat{N}_\mathrm{fluct}\rangle \approx 0.3$. This corresponds to a photon flux of about $10^3$~sec$^{-1}$. For the example of an incoming single photon, incoherent photons will make a relatively small contribution for photon pulses with a bandwidth larger than $10$~kHz.

\begin{figure}[t]
\centerline{\includegraphics[height=4.5cm]{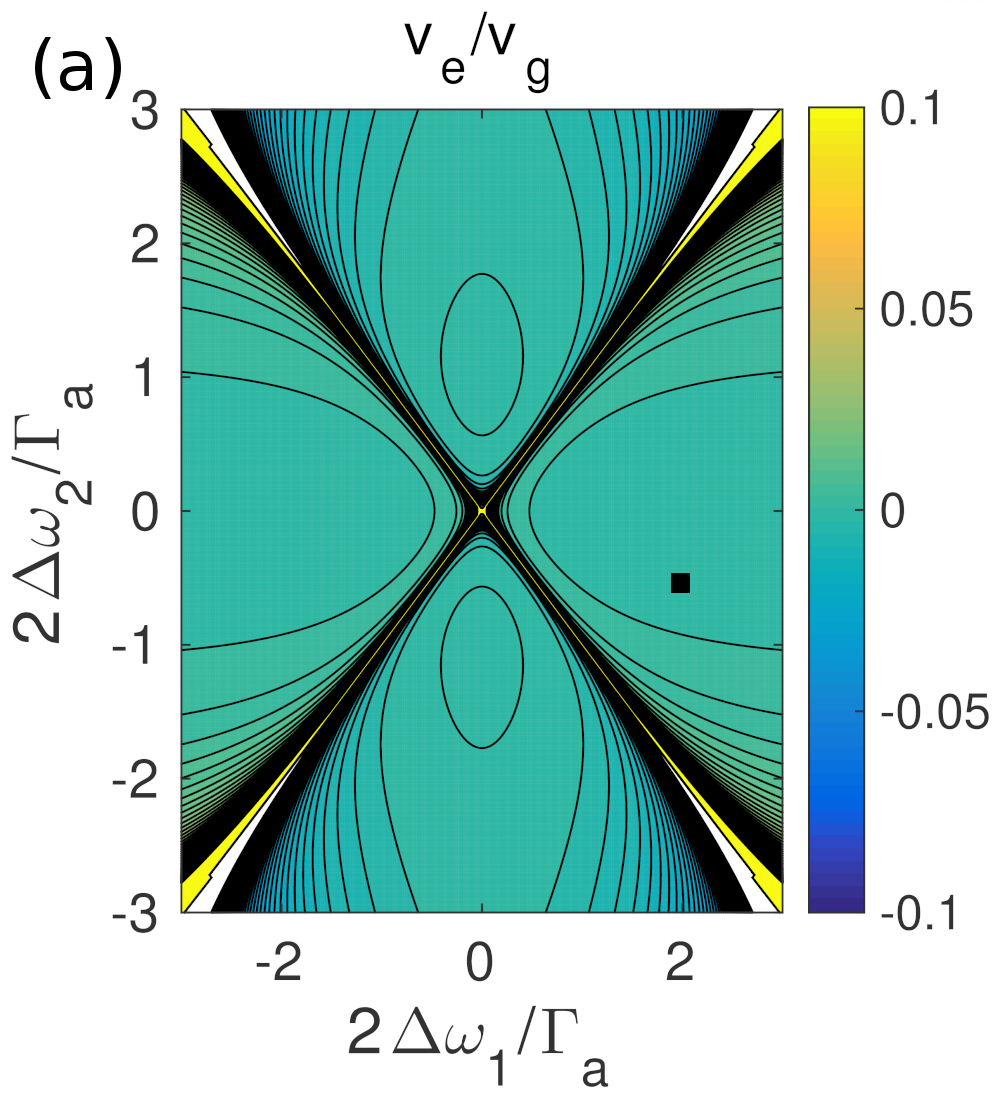}\includegraphics[height=4.5cm]{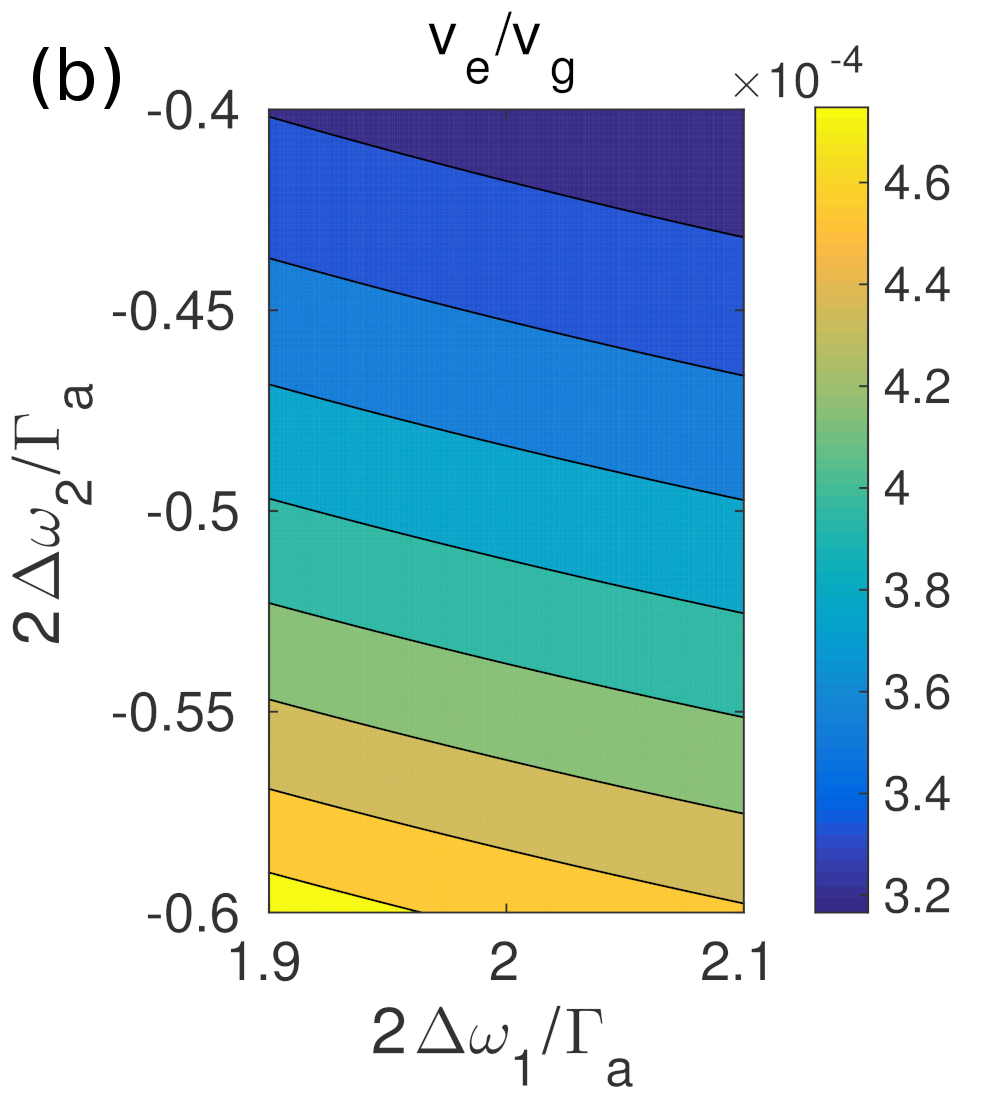}}
\centerline{\includegraphics[height=4.5cm]{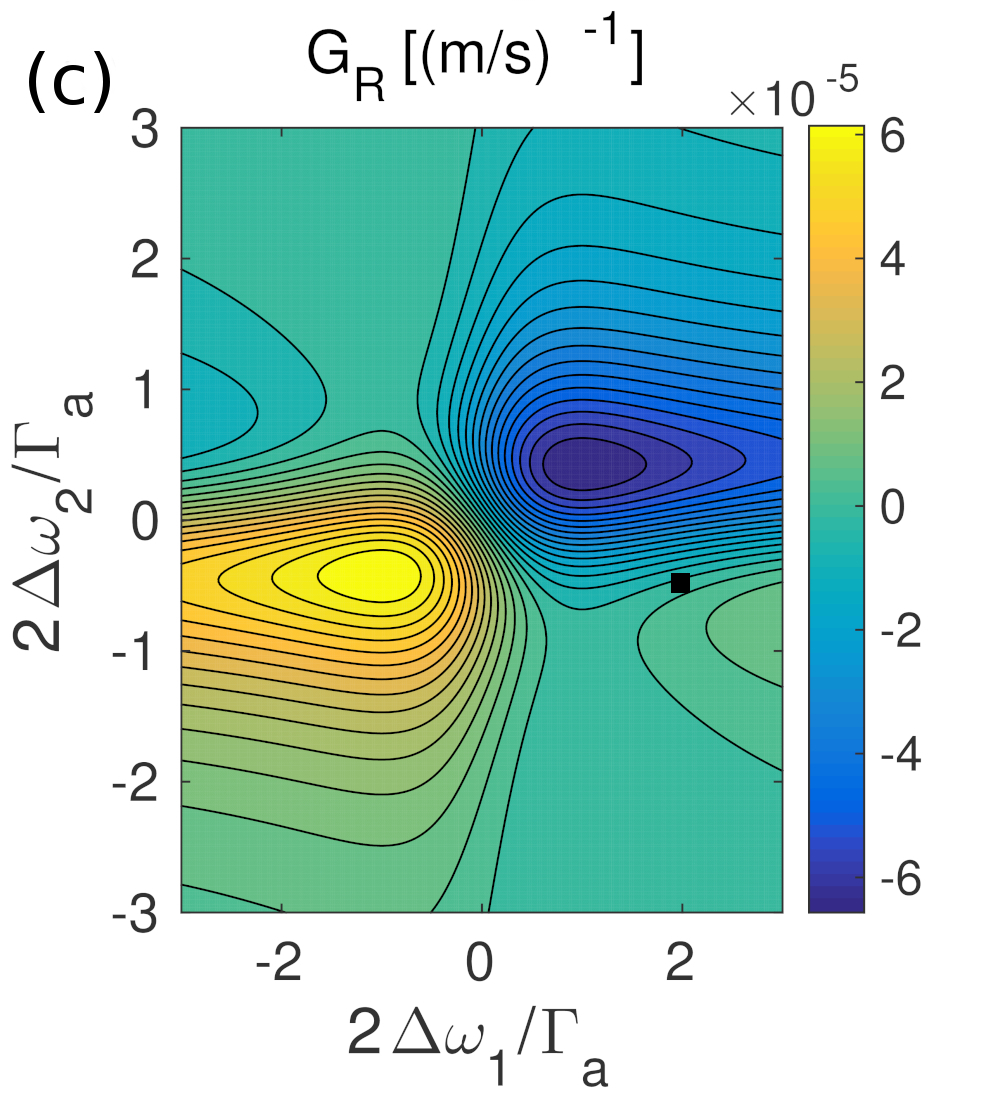}\includegraphics[height=4.5cm]{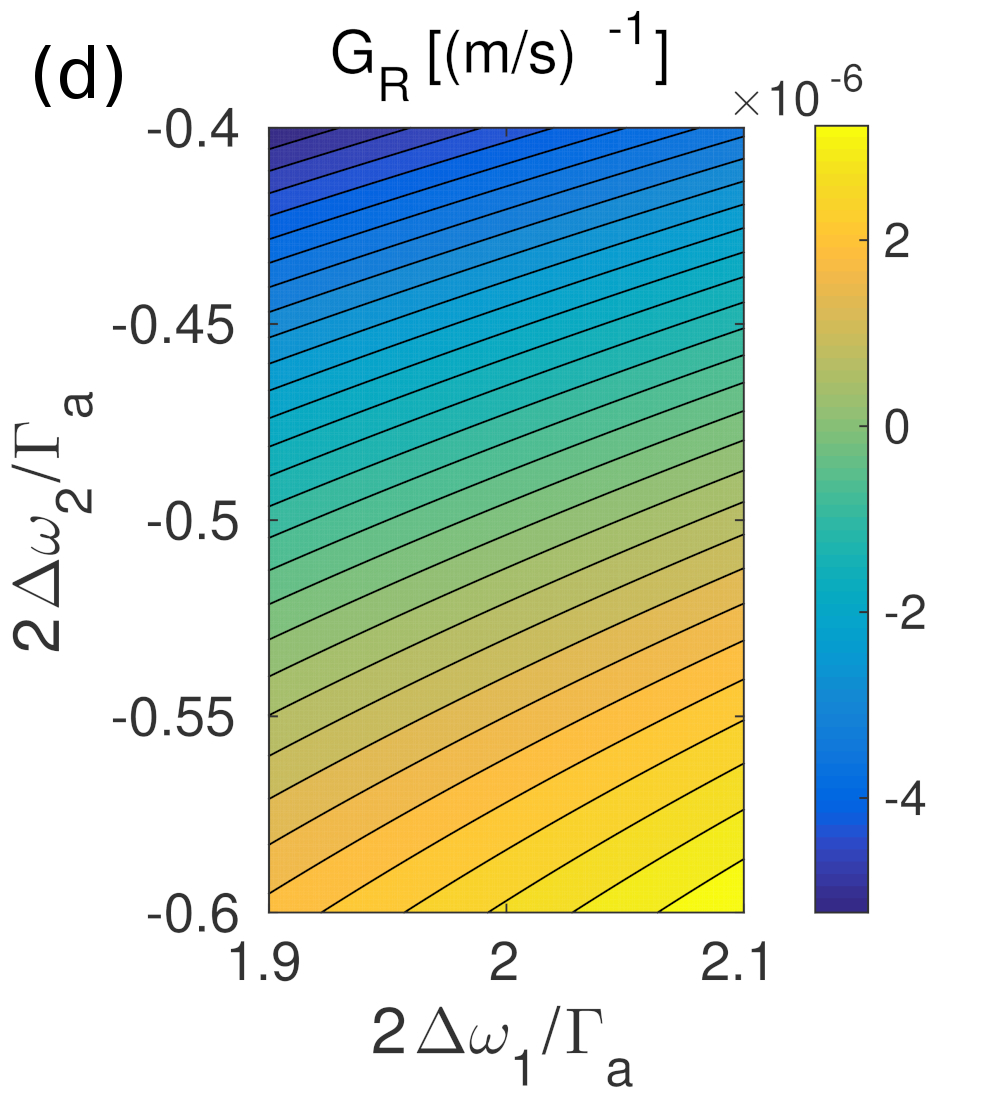}}
\caption{(a) The reduction in group velocity $v_e/v_g$ versus the detunings
  $b_1=\frac{2\Delta\omega_1}{\Gamma_a}$ and
  $b_2=\frac{2\Delta\omega_2}{\Gamma_a}$. (b) Zoom around the working point
  $\left(b_1=2,b_2=-\frac{1}{2}\right)$ marked in (a). (c) The gradient of the gain $G_R=\frac{\partial
  G}{\partial\omega_s}$. (d) Zoom around
  the point $\left(b_1=2,b_2=-\frac{1}{2}\right)$ marked in (c).}
\label{VelocityEff}
\end{figure}

Now we have $\psi(x,t)=e^{i(Kx-\omega_st)}e^{-Gx}\psi_{in}(x-v_gt)$, where
$K=k_s-\kappa$. The effective group velocity is defined by
$\frac{1}{v_e}=\frac{dK}{d\omega_s}$. Our goal now is to identify parameter regimes exhibiting a small group velocity $v_e$ and, at the same time, vanishing gain $G$ in a sufficiently broad bandwidth, that is with a small gradient $G_R=\frac{\partial
  G}{\partial\omega_s}$. The control parameters are the intensities $\mathcal{E}_i$ and detunings $\Delta\omega_i$ of the pump fields. It will be convenient to use dimensionless detunings $b_i=\Delta\omega_i/\frac{\Gamma_a}{2}$. For given detunings a vanishing gain $G=0$ is achieved for an intensity ratio of $|{\cal
  E}_1|^2/|{\cal E}_2|^2=\frac{1+b_1^2}{1+b_2^2}$ which we assume to be fulfilled in the following. Using the same numbers as above and $\Gamma_a=10^8$~Hz corresponding to a mechanical quality factor of $Q_a=10^3$ we show the reduction in group velocity$\frac{v_e}{v_g}$ in Fig.~\ref{VelocityEff}.a and the gradient of the gain $G_R$ in Fig.~\ref{VelocityEff}.c versus the detunings. A convenient working point is found at $\left(b_1=2,b_2=-\frac{1}{2}\right)$ which is shown in more detail in Figs.~\ref{VelocityEff}.b and \ref{VelocityEff}.d. At this point we have $|{\cal
  E}_1|^2=4|{\cal E}_2|^2$, and $\Delta\omega_1=\Gamma_a$,
$\Delta\omega_2=-\frac{\Gamma_a}{4}$. The reduction in group velocity is
$\frac{v_e}{v_g}\approx\frac{\Gamma_a^2}{4f_a^2|{\cal E}|^2}$, and the above numbers yields $\frac{v_e}{v_g}\approx 3.7\times 10^{-4}$ which corresponds to what we have assumed above.

In conclusion, we predict that quantum nonlinear optics, slow light without gain/loss, and nonlinear phase shifts are possible in nanoscale waveguides exploiting SBS. Even though we considered here the most simple geometry of a cylindrical fibre, it is clear that coupling strengths and quality factors can be further optimized using different geometries of the nanostructure, see \cite{Russell2006} and \cite{Rakich2016} for examples. The present results provide encouraging evidence for the realization of many-body physics with strongly interacting photons and the implementation of deterministic quantum gates for photons in continuum optomechancis.

\paragraph*{Acknowledgments} This work was funded by the European Commission (FP7-Programme) through iQUOEMS (Grant Agreement No. 323924). We acknowledge support by DFG through QUEST. We thank Raphael Van Laer for fruitful discussions.

\onecolumngrid

\newpage

\section*{\large Supplemental Materials: \\
Nonlinear Quantum Optics in Optomechanical Nanoscale Waveguides}

\section*{Hashem Zoubi and Klemens Hammerer}

\section{Photon correlations mediated by vibrational modes}

Starting from the Hamiltonian, which was derived in \cite{Zoubi2016},
\begin{align}
H&=\sum_{\alpha}\int
dx\ \omega_\alpha\psi_{\alpha}^{\dagger}(x)\psi_{\alpha}(x)+\Omega_v\int dx\ Q^{\dagger}(x)Q(x) \nonumber \\
&-i v_e\sum_{\alpha}\int
dx\ \psi_{\alpha}^{\dagger}(x)\frac{\partial\psi_{\alpha}(x)}{\partial
  x}+\sqrt{L}f_v\int dx\ \left(Q^{\dagger}(x)\psi_d^{\dagger}(x)\psi_u(x)+h.c.\right)\nonumber, \nonumber
\end{align}
using
$\psi_{\alpha}(x,t)\rightarrow\psi_{\alpha}(x,t)\ e^{ik_{\alpha}x}$ and
$Q_{v}(x,t)\rightarrow Q_{v}(x,t)\ e^{iq_{v}x}$, we obtain Hamiltonian (1)
of the letter, which yields the equations of motion for the field operators
\begin{align}
\left(\frac{\partial}{\partial t}+v_e\frac{\partial}{\partial
  x}\right)\psi_u(x,t)&=-i\omega_u\ \psi_u(x,t)-if_v\sqrt{L}\ Q_v(x,t)\psi_d(x,t)e^{-i\Delta qx}, \nonumber \\
\left(\frac{\partial}{\partial t}+v_e\frac{\partial}{\partial
  x}\right)\psi_d(x,t)&=-i\omega_d\ \psi_d(x,t)-if_v\sqrt{L}\ Q_v^{\dagger}(x,t)\psi_u(x,t)e^{i\Delta qx}, \nonumber \\
\left(\frac{\partial}{\partial
  t}+\frac{\Gamma_v}{2}\right)Q_{v}(x,t)&=-i\Omega_v\ Q_v(x,t)-if_v\sqrt{L}\ \psi_d^{\dagger}(x,t)\psi_u(x,t)e^{i\Delta qx}-{\cal
  F}(x,t), \nonumber
\end{align}
where $\Delta
q=k_u-k_d-q_v$. Using now
$\psi_{\alpha}(x,t)\rightarrow\psi_{\alpha}(x,t)\ e^{-i\omega_{\alpha}t}$,
$Q_{v}(x,t)\rightarrow Q_{v}(x,t)\ e^{-i\Omega_{v}t}$, and ${\cal
  F}(x,t)\rightarrow {\cal F}(x,t)\ e^{-i\Omega_{v}t}$, we obtain
\begin{align}
\left(\frac{\partial}{\partial t}+v_e\frac{\partial}{\partial
  x}\right)\psi_u(x,t)&=-if_v\sqrt{L}\ Q_v(x,t)\psi_d(x,t)\ e^{i(\Delta\Omega t-\Delta qx)}, \nonumber \\
\left(\frac{\partial}{\partial t}+v_e\frac{\partial}{\partial
  x}\right)\psi_d(x,t)&=-if_v\sqrt{L}\   Q_v^{\dagger}(x,t)\psi_u(x,t)\ e^{-i(\Delta\Omega t-\Delta qx)}, \nonumber \\
\left(\frac{\partial}{\partial
  t}+\frac{\Gamma_v}{2}\right)Q_{v}(x,t)&=-if_v\sqrt{L}\ \psi_d^{\dagger}(x,t)\psi_u(x,t)\ e^{-i(\Delta\Omega
  t-\Delta qx)}-{\cal F}(x,t), \nonumber
\end{align}
where $\Delta\Omega=\omega_u-\omega_d-\Omega_v$. The above system of equations corresponds to Eqs. (2-3) in the main text. We now apply the adiabatic elimination of the
phonon operators, which is applicable in the off resonant limit where
$\Delta\Omega>\Gamma_v,f_v$. Formal integration of the phonon equation gives
\begin{align}
Q(x,t)&=-i\sqrt{L}f_v\int_{0}^{t}dt'\ \psi_d^{\dagger}(x,t')\psi_u(x,t')\ e^{-i(\Delta\Omega
  t'-\Delta qx)}e^{-\Gamma_v(t-t')/2} \nonumber \\
&+Q(x,0)e^{-\Gamma_v t/2}-\int_{0}^t dt'\ {\cal F}(x,t')e^{-\frac{\Gamma_v}{2}(t-t')}. \nonumber
\end{align}
We neglect the initial value term of the phonon operator. Substitution in the photon equations yields
\begin{align}
\left(\frac{\partial}{\partial t}+v_e\frac{\partial}{\partial
  x}\right)\psi_u(x,t)&=-f_v^2L\int_{0}^{t}dt'\ \psi_d^{\dagger}(x,t')\psi_u(x,t')\psi_d(x,t)\ e^{-i\Delta\Omega
  (t'-t)}e^{-\Gamma_v(t-t')/2} \nonumber \\
&+if_v\sqrt{L}\ \psi_d(x,t)e^{i(\Delta\Omega t-\Delta
  qx)}\int_{0}^t dt'\ {\cal F}(x,t')e^{-\frac{\Gamma_v}{2}(t-t')}, \nonumber \\
\left(\frac{\partial}{\partial t}+v_e\frac{\partial}{\partial
  x}\right)\psi_d(x,t)&=f_v^2L\int_{0}^{t}dt'\ \psi_u^{\dagger}(x,t')\psi_d(x,t')\psi_u(x,t)\ e^{i\Delta\Omega (t'-t)}e^{-\Gamma_v(t-t')/2} \nonumber \\
&+if_v\sqrt{L}\ \psi_u(x,t)e^{-i(\Delta\Omega t-\Delta
  qx)}\int_{0}^t dt'\ {\cal F}^{\dagger}(x,t')e^{-\frac{\Gamma_v}{2}(t-t')}. \nonumber
\end{align}
Now we apply an approximation by taking the operators out of the integral,
which is allowed in the limit $\Delta\Omega> f_v$, to get
\begin{align}
\left(\frac{\partial}{\partial t}+v_e\frac{\partial}{\partial
  x}\right)\psi_u(x,t)&\approx -Lf_v^2\ \hat{N}_d(x,t)\psi_u(x,t)\int_{0}^{t}dt'\ e^{-i\Delta\Omega
  (t'-t)}e^{-(t-t')\Gamma_v/2}+iU(x,t)\ \psi_d(x,t), \nonumber \\
\left(\frac{\partial}{\partial t}+v_e\frac{\partial}{\partial
  x}\right)\psi_d(x,t)&\approx Lf_v^2\ \hat{N}_u(x,t)\psi_d(x,t)\int_{0}^{t}dt'\ e^{i\Delta\Omega (t'-t)}e^{-(t-t')\Gamma_v/2}+iU^{\dagger}(x,t)\ \psi_u(x,t), \nonumber
\end{align}
where we defined the density operator by
$\hat{N}_{\alpha}(x,t)=\psi_{\alpha}^{\dagger}(x,t)\psi_{\alpha}(x,t)$. Moreover,
we used
\begin{align}
U(x,t)=f_v\sqrt{L}\ e^{i(\Delta\Omega t-\Delta
  qx)}\int_{0}^t dt'\ {\cal F}(x,t')e^{-\frac{\Gamma_v}{2}(t-t')}. \nonumber
\end{align}
The time integration yields
\begin{align}
\left(\frac{\partial}{\partial t}+v_e\frac{\partial}{\partial
  x}\right)\psi_u(x,t)&\approx-\frac{Lf_v^2}{\Gamma_v/2-i\Delta\Omega}\ \hat{N}_d(x,t)\psi_u(x,t)+iU(x,t)\ \psi_d(x,t), \nonumber \\
\left(\frac{\partial}{\partial t}+v_e\frac{\partial}{\partial
  x}\right)\psi_d(x,t)&\approx\frac{Lf_v^2}{\Gamma_v/2+i\Delta\Omega}\ \hat{N}_u(x,t)\psi_d(x,t)+iU^{\dagger}(x,t)\ \psi_u(x,t). \nonumber
\end{align}

\subsection{Conserved Number of Photons}

We show now that the total density of signal photons, that is $\hat{N}=\hat{N}_u+\hat{N}_d$, is
conserved. We drop the Langevin term in this part and consider it later. Direct calculations give
\begin{align}
\left(\frac{\partial}{\partial t}+v_e\frac{\partial}{\partial
  x}\right)\hat{N}_u(x,t)&=-\frac{Lf_v^2\Gamma_v}{\Gamma_v^2/4+\Delta\Omega^2}\ \hat{N}_u(x,t)\hat{N}_d(x,t), \nonumber \\
\left(\frac{\partial}{\partial t}+v_e\frac{\partial}{\partial
  x}\right)\hat{N}_d(x,t)&=\frac{Lf_v^2\Gamma_v}{\Gamma_v^2/4+\Delta\Omega^2}\ \hat{N}_u(x,t)\hat{N}_d(x,t), \nonumber
\end{align}
which yields $\left(\frac{\partial}{\partial t}+v_e\frac{\partial}{\partial
  x}\right)\hat{N}(x,t)=0$. Using the change of variables $\xi=x-v_et$ and $\eta=v_et$, gives
$\frac{\partial}{\partial t}+v_e\frac{\partial}{\partial
  x}=v_e\frac{\partial}{\partial \eta}$ and $\frac{\partial}{\partial
  \eta}\hat{N}(\xi,\eta)=0$, and hence $\hat{N}(\xi)$ is conserved. Here we obtain
\begin{align}
\frac{\partial}{\partial
  \eta}\hat{N}_u(\xi,\eta)=-V\ \hat{N}_d(\xi,\eta)\hat{N}_u(\xi,\eta),\ \ \ \frac{\partial}{\partial \eta}\hat{N}_d(\xi,\eta)=V\ \hat{N}_u(\xi,\eta)\hat{N}_d(\xi,\eta), \nonumber
\end{align}
where $V=\frac{Lf_v^2\Gamma_v}{v_e(\Gamma_v^2/4+\Delta\Omega^2)}$. Using
$\hat{N}(\xi)=\hat{N}_u(\xi,\eta)+\hat{N}_d(\xi,\eta)$ gives the two Riccati
equations (4) of the letter.

\subsection{Thermal Fluctuations}

In the letter it was shown
that $\hat{N}_u$ and $\hat{N}_d$ are conserved in the limit $\Delta\Omega>\Gamma_v$. Hence, we get
\begin{align}
\left(\frac{\partial}{\partial t}+v_e\frac{\partial}{\partial
  x}\right)\psi_u(x,t)&\approx-i\frac{Lf_v^2}{\Delta\Omega}\ \hat{N}_d(x,t)\psi_u(x,t)+iU(x,t)\ \psi_d(x,t), \nonumber \\
\left(\frac{\partial}{\partial t}+v_e\frac{\partial}{\partial
  x}\right)\psi_d(x,t)&\approx-i\frac{Lf_v^2}{\Delta\Omega}\ \hat{N}_u(x,t)\psi_d(x,t)+iU^{\dagger}(x,t)\ \psi_u(x,t). \nonumber
\end{align}
We calculate here the contribution of the
Langevin fluctuations. Applying the change of variables $\xi=x-v_et$ and $\eta=x$, where
$\frac{\partial}{\partial t}=-v_e\frac{\partial}{\partial \xi}$ and
$\frac{\partial}{\partial x}=\frac{\partial}{\partial
  \xi}+\frac{\partial}{\partial \eta}$, and then $\frac{\partial}{\partial
  t}+v_e\frac{\partial}{\partial x}=v_g\frac{\partial}{\partial \eta}$, we get
\begin{align}
\frac{\partial}{\partial\eta}\psi_u(\xi,\eta)&\approx-i\frac{Lf_v^2}{v_e\Delta\Omega}\ \hat{N}_d(\xi)\psi_u(\xi,\eta)+\frac{i}{v_e}U(\xi,\eta)\ \psi_d(\xi,\eta), \nonumber \\
\frac{\partial}{\partial\eta}\psi_d(\xi,\eta)&\approx-i\frac{Lf_v^2}{v_e\Delta\Omega}\ \hat{N}_u(\xi)\psi_d(\xi,\eta)+\frac{i}{v_e}U^{\dagger}(\xi,\eta)\ \psi_u(\xi,\eta). \nonumber
\end{align}
Formal integration gives
\begin{align}
\psi_u(\xi,\eta)=\psi_u^\mathrm{in}(\xi)e^{-i\frac{Lf_v^2}{v_e\Delta\Omega}\ \hat{N}_d(\xi)\ \eta}+\frac{i}{v_e}\int_0^{\eta}d\eta'\ U(\xi,\eta')\ \psi_d(\xi,\eta')e^{-i\frac{Lf_v^2}{v_e\Delta\Omega}\ \hat{N}_d(\xi)(\eta-\eta')},
\nonumber \\
\psi_d(\xi,\eta)=\psi_d^\mathrm{in}(\xi)e^{-i\frac{Lf_v^2}{v_e\Delta\Omega}\ \hat{N}_u(\xi)\ \eta}+\frac{i}{v_e}\int_0^{\eta}d\eta'\ U^{\dagger}(\xi,\eta')\ \psi_u(\xi,\eta')e^{-i\frac{Lf_v^2}{v_e\Delta\Omega}\ \hat{N}_u(\xi)(\eta-\eta')}, \nonumber
\end{align}
where
$\psi_{\alpha}^{in}(\xi)=\psi_{\alpha}(\xi,\eta=0)$. Changing back into
$(x,t)$ space we get Equ. (5) of the letter. The average numbers of photons are
\begin{align}
&\langle\psi^{\dagger}_u(x,t)\psi_u(x,t)\rangle=\langle\psi^{\mathrm{in}\dagger}_u(x-v_et)\psi_u^\mathrm{in}(x-v_et)\rangle\nonumber
\\
&+\frac{1}{v_e^2}\int_0^{x}dx'dx''\ \langle\psi^{\dagger}_d(x',t)\psi_d(x'',t)\rangle\langle U^{\dagger}(x',t)U(x'',t)\rangle
e^{i\frac{Lf_v^2}{v_e\Delta\Omega}\ N_d(x-v_et)(x-x')}e^{-i\frac{Lf_v^2}{v_e\Delta\Omega}\ N_d(x-v_et)(x-x'')}\nonumber \\
&\langle\psi^{\dagger}_d(x,t)\psi_d(x,t)\rangle=\langle\psi^{\mathrm{in}\dagger}_d(x-v_et)\psi_d^\mathrm{in}(x-v_et)\rangle
\nonumber \\
&+\frac{1}{v_e^2}\int_0^{x}dx'dx''\ \langle\psi^{\dagger}_u(x',t)\psi_u(x'',t)\rangle\langle U(x',t)U^{\dagger}(x'',t)\rangle
e^{i\frac{Lf_v^2}{v_e\Delta\Omega}\ N_u(x-v_et)(x-x')}e^{-i\frac{Lf_v^2}{v_e\Delta\Omega}\ N_u(x-v_et)(x-x'')}, \nonumber
\end{align}
where $N_{\alpha}=\langle\hat{N}_{\alpha}\rangle$. Using
\begin{align}
\langle
{\cal F}^{\dagger}(x',t'){\cal F}(x'',t'')\rangle&=\Gamma_a\bar{n}_v\delta(t'-t'')\delta(x'-x''),
\nonumber \\
\langle
{\cal F}(x',t'){\cal F}^{\dagger}(x'',t'')\rangle&=\Gamma_a(\bar{n}_v+1)\delta(t'-t'')\delta(x'-x''), \nonumber
\end{align}
where $\bar{n}_v$ is the average number of thermal phonons, we get
\begin{align}
\langle U^{\dagger}(x',t)U(x'',t)\rangle&=Lf_v^2\bar{n}_v\delta(x'-x'')\left(1-e^{-\Gamma_vt}\right),
\nonumber \\
\langle U(x',t)U^{\dagger}(x'',t)\rangle&=Lf_v^2(1+\bar{n}_v)\delta(x'-x'')\left(1-e^{-\Gamma_vt}\right). \nonumber
\end{align}
We obtain
\begin{align}
\langle\psi^{\dagger}_u(x,t)\psi_u(x,t)\rangle&=\langle\psi^{\mathrm{in}\dagger}_u(x-v_et)\psi_u^\mathrm{in}(x-v_et)\rangle+\frac{Lf_v^2}{v_e^2}\bar{n}_v\left(1-e^{-\Gamma_vt}\right)\int_0^{x}dx'\ \langle\psi^{\dagger}_d(x',t)\psi_d(x',t)\rangle,
\nonumber \\
\langle\psi^{\dagger}_d(x,t)\psi_d(x,t)\rangle&=\langle\psi^{\mathrm{in}\dagger}_d(x-v_et)\psi_d^\mathrm{in}(x-v_et)\rangle+\frac{Lf_v^2}{v_e^2}(1+\bar{n}_v)\left(1-e^{-\Gamma_vt}\right)\int_0^{x}dx'\ \langle\psi^{\dagger}_u(x',t)\psi_u(x',t)\rangle, \nonumber
\end{align}
which yields at the waveguide output
\begin{align}
N_u=N_u^\mathrm{in}+\frac{L^2f_v^2}{v_e^2}\left(1-e^{-\Gamma_vL/v_e}\right)\bar{n}_vN_d,\ \ \ N_d=N_d^\mathrm{in}+\frac{L^2f_v^2}{v_e^2}\left(1-e^{-\Gamma_vL/v_e}\right)(1+\bar{n}_v)N_u. \nonumber
\end{align}

\section{Photon delay and gain via SBS involving acoustic phonons}

The real-space Hamiltonian is given in equation (6) of the letter. The photon dispersion has the form
$\omega_k\approx\omega_0\pm v_gk$, and all photon fields are taken in a
rotating frame of their respective central frequency $\omega_0$. We get the equations of motion for the field operators
\begin{align}
\left(\frac{\partial}{\partial t}+v_g\frac{\partial}{\partial
  x}\right)\psi(x,t)&=-if^a_1{\cal E}_{1}(x,t)\ Q^a_1(x,t)-i f^a_2{\cal E}_{2}(x,t)\ Q_2^{a\dagger}(x,t), \nonumber \\
\left(\frac{\partial}{\partial t}+v_a\frac{\partial}{\partial
  x}+\frac{\Gamma_a}{2}\right)Q^a_{1}(x,t)&=-i f_1^{a}{\cal E}_{1}^{\ast}(x,t)\ \psi(x,t)-{\cal F}_1(x,t),
\nonumber \\
\left(\frac{\partial}{\partial t}-v_a\frac{\partial}{\partial
  x}+\frac{\Gamma_a}{2}\right)Q^a_{2}(x,t)&=-i f^a_2{\cal E}_{2}(x,t)\ \psi^{\dagger}(x,t)-{\cal F}_2(x,t). \nonumber
\end{align}
We use ${\cal E}_{\alpha}(x,t)\rightarrow{\cal
  E}_{\alpha}(x,t)\ e^{-i(k_{\alpha}x+\omega_{\alpha}t)}$,
$\psi(x,t)\rightarrow\psi(x,t)\ e^{i(k_sx-\omega_st)}$,
$Q_{1}^a(x,t)\rightarrow Q_{1}^a(x,t)\ e^{i(q_{1}x-\Omega_{1}t)}$, and
$Q_{2}^a(x,t)\rightarrow Q_{2}^a(x,t)\ e^{-i(q_{2}x+\Omega_{2}t)}$, with
$(\alpha=1,2)$, where
$\omega_s=v_gk_s$, $\omega_{\alpha}=v_gk_{\alpha}$, and
$\Omega_{\alpha}=v_aq_{\alpha}$. Moreover we define ${\cal
  F}_{1}(x,t)\rightarrow{\cal F}_{1}(x,t)\ e^{i(q_{1}x-\Omega_{1}t)}$, and ${\cal
  F}_{2}(x,t)\rightarrow{\cal F}_{2}(x,t)\ e^{-i(q_{2}x+\Omega_{2}t)}$. We have now
\begin{align}
\left(\frac{\partial}{\partial t}+v_g\frac{\partial}{\partial
  x}\right)\psi(x,t)&=-i f^a_1{\cal
  E}_{1}(x,t)\ Q_1^a(x,t)\ e^{i(\Delta\omega_1t-\Delta k_1x)}-i f^a_2{\cal
  E}_{2}(x,t)\ Q_2^{a\dagger}(x,t)\ e^{-i(\Delta\omega_2t+\Delta k_2x)}, \nonumber \\
\left(\frac{\partial}{\partial t}+v_a\frac{\partial}{\partial
  x}+\frac{\Gamma_a}{2}\right)Q_{1}^a(x,t)&=-i f_1^{a}{\cal E}_{1}^{\ast}(x,t)\ \psi(x,t)\ e^{-i(\Delta\omega_1t-\Delta k_1x)}-{\cal F}_{1}(x,t),
\nonumber \\
\left(\frac{\partial}{\partial t}-v_a\frac{\partial}{\partial
  x}+\frac{\Gamma_a}{2}\right)Q_{2}^a(x,t)&=-i f^a_2{\cal E}_{2}(x,t)\ \psi^{\dagger}(x,t)\ e^{-i(\Delta\omega_2t+\Delta k_2x)}-{\cal F}_{2}(x,t), \nonumber
\end{align}
where $\Delta\omega_1=\omega_s-\omega_1-\Omega_1$,
$\Delta\omega_2=\omega_2-\omega_s-\Omega_2$, $\Delta k_1=k_s+k_1-q_1$ and $\Delta k_2=k_2+k_s-q_2$. The above system of equations corresponds to Eqs. (7-8) in the main text. For acoustic phonons it
is a good approximation to neglect the $v_a\frac{\partial}{\partial
  x}$ terms, as the sound velocity is much smaller than the light group
velocity, then
\begin{align}
\left(\frac{\partial}{\partial
  t}+\frac{\Gamma_a}{2}\right)Q_{1}^a(x,t)&\approx-i f_1^{a}{\cal E}_{1}^{\ast}(x,t)\ \psi(x,t)\ e^{-i(\Delta\omega_1t-\Delta k_1x)}-{\cal F}_{1}(x,t),
\nonumber \\
\left(\frac{\partial}{\partial
  t}+\frac{\Gamma_a}{2}\right)Q_{2}^a(x,t)&\approx-i f^a_2{\cal E}_{2}(x,t)\ \psi^{\dagger}(x,t)\ e^{-i(\Delta\omega_2t+\Delta k_2x)}-{\cal F}_{2}(x,t). \nonumber
\end{align}

Formal integration of the phonon operators lead to
\begin{align}
Q_1^a(x,t)&=-i f_1^{a}\int_{0}^t dt'\ {\cal
  E}_{1}^{\ast}(x,t')\ \psi(x,t')\ e^{-i(\Delta\omega_1t'-\Delta
  k_1x)}e^{-\frac{\Gamma_a}{2}(t-t')} \nonumber \\
&+Q_1(x,0)e^{-\Gamma_a t/2}-\int_{0}^t dt'\ {\cal F}_{1}(x,t')e^{-\frac{\Gamma_a}{2}(t-t')},
\nonumber \\
Q_2^a(x,t)&=-i f^a_2\int_{0}^t dt'\ {\cal
  E}_{2}(x,t')\ \psi^{\dagger}(x,t')\ e^{-i(\Delta\omega_2t'+\Delta
  k_2x)}e^{-\frac{\Gamma_a}{2}(t-t')} \nonumber \\
&+Q_2(x,0)e^{-\Gamma_a t/2}-\int_{0}^t dt'\ {\cal F}_{2}(x,t')e^{-\frac{\Gamma_a}{2}(t-t')}.
\nonumber
\end{align}
In the following we neglect the initial value terms of the phonon operators. As an approximation we take the signal operator and the pump field out of the integral to get
\begin{align}
Q_1^a(x,t)&\approx-i f_1^{a}\ {\cal E}_{1}^{\ast}(x,t)\ \psi(x,t)\int_{0}^t dt'\ e^{-i(\Delta\omega_1t'-\Delta k_1x)}e^{-\frac{\Gamma_a}{2}(t-t')}-\int_{0}^t dt'\ {\cal F}_{1}(x,t')e^{-\frac{\Gamma_a}{2}(t-t')},
\nonumber \\
Q_2^a(x,t)&\approx-i f^a_2\ {\cal E}_{2}(x,t)\ \psi^{\dagger}(x,t)\int_{0}^t dt'\ e^{-i(\Delta\omega_2t'+\Delta k_2x)}e^{-\frac{\Gamma_a}{2}(t-t')}-\int_{0}^t dt'\ {\cal F}_{2}(x,t')e^{-\frac{\Gamma_a}{2}(t-t')}. \nonumber
\end{align}
This approximation is an iterative solution in terms of the small photon-phonon coupling
parameter. Substitution in the signal operator equation yields
\begin{align}
\left(\frac{\partial}{\partial t}+v_g\frac{\partial}{\partial
  x}\right)\psi(x,t)&=-f^{a2}_1|{\cal
  E}_{1}(x,t)|^2\int_{0}^t dt'\ e^{-i\Delta\omega_1(t'-t)}e^{-\frac{\Gamma_a}{2}(t-t')}\ \psi(x,t) \nonumber \\
&+f^{a2}_2|{\cal
  E}_{2}(x,t)|^2\int_{0}^t
dt'\ e^{i\Delta\omega_2(t'-t)}e^{-\frac{\Gamma_a}{2}(t-t')}\ \psi(x,t) \nonumber \\
&+i f^a_1{\cal
  E}_{1}(x,t)\ e^{i(\Delta\omega_1t-\Delta k_1x)}\int_{0}^t dt'\ {\cal F}_{1}(x,t')e^{-\frac{\Gamma_a}{2}(t-t')} \nonumber \\
&+i f^a_2{\cal
  E}_{2}(x,t)\ e^{-i(\Delta\omega_2t+\Delta k_2x)}\int_{0}^t dt'\ {\cal F}_{2}^{\dagger}(x,t')e^{-\frac{\Gamma_a}{2}(t-t')}. \nonumber
  \end{align}
Time integration gives
\begin{align}
\left(\frac{\partial}{\partial t}+v_g\frac{\partial}{\partial
  x}\right)\psi(x,t)&\approx f_a^2\left\{-\frac{|{\cal
  E}_{1}(x,t)|^2}{\frac{\Gamma_a}{2}-i\Delta\omega_1}+\frac{|{\cal
  E}_{2}(x,t)|^2}{\frac{\Gamma_a}{2}+i\Delta\omega_2}\right\}\psi(x,t)
\nonumber \\
&+if_a\left\{{\cal E}_{1}(x,t)e^{-i\Delta k_1x}W_1(x,t)+{\cal E}_{2}(x,t)e^{-i\Delta k_2x}W_2^{\dagger}(x,t)\right\}, \nonumber
\end{align}
where we assume that $f^a_1=f^a_2\equiv f_a$. We defined
\begin{align}
W_i(x,t)=e^{i\Delta\omega_it}\int_{0}^t dt'\ {\cal F}_{i}(x,t')e^{-\frac{\Gamma_a}{2}(t-t')}. \nonumber
\end{align}
We can write
\begin{align}
\left(\frac{\partial}{\partial t}+v_g\frac{\partial}{\partial
  x}\right)\psi(x,t)=-v_g(G+i\kappa)\psi(x,t)+if_a\left\{{\cal E}_{1}e^{-i\Delta k_1x}W_1(x,t)+{\cal E}_{2}e^{-i\Delta k_2x}W_2^{\dagger}(x,t)\right\}, \nonumber
\end{align}
where $G$ and $\kappa$ are defined in the letter. The pump fields are taken to be constants.

\subsection{Thermal Fluctuations}

Applying the previous change of variables $\xi=x-v_gt$ and $\eta=x$, we get
\begin{align}
\frac{\partial}{\partial\eta}\psi(\xi,\eta)=-(G+i\kappa)\psi(\xi,\eta)+i\frac{f_a}{v_g}\left\{{\cal E}_{1}e^{-i\Delta k_1\eta}W_1(\xi,\eta)+{\cal E}_{2}e^{-i\Delta k_2\eta}W_2^{\dagger}(\xi,\eta)\right\}, \nonumber
\end{align}
with the solution
\begin{align}
\psi(\xi,\eta)=e^{-(G+i\kappa)\eta}\psi_{in}(\xi)+i\frac{f_a}{v_g}\int_0^{\eta}d\eta'\left\{{\cal E}_{1}e^{-i\Delta k_1\eta'}W_1(\xi,\eta')+{\cal E}_{2}e^{-i\Delta k_2\eta'}W_2^{\dagger}(\xi,\eta')\right\}e^{(G+i\kappa)(\eta'-\eta)}, \nonumber
\end{align}
where $\psi_{in}(\xi)=\psi(\xi,\eta=0)$. Back into $(x,t)$
variables we obtain equation (9) of the letter. The average density of photons is
\begin{align}
\langle\psi^{\dagger}(x,t)\psi(x,t)\rangle&=\langle\psi_\mathrm{in}^{\dagger}(x-v_gt)\psi_\mathrm{in}(x-v_gt)\rangle
e^{-2Gx}+\frac{f_a^2}{v_g^2}\int_0^xdx'dx''e^{(G-i\kappa)(x'-x)}e^{(G+i\kappa)(x''-x)}
\nonumber \\
&\times\left\{|{\cal E}_{1}|^2e^{i\Delta k_1(x'-x'')}\langle
W_1^{\dagger}(x',t)W_1(x'',t)\rangle+|{\cal E}_{2}|^2e^{i\Delta k_2(x'-x'')}\langle
W_2(x',t)W_2^{\dagger}(x'',t)\rangle \right. \nonumber \\
&+\left. {\cal E}_{1}^{\ast}{\cal E}_{2}e^{i(\Delta k_1x'-\Delta k_2x'')}\langle
W_1^{\dagger}(x',t)W_2^{\dagger}(x'',t)\rangle+{\cal E}_{2}^{\ast}{\cal
  E}_{1}e^{i(\Delta k_2x'-\Delta k_1x'')} \langle W_2(x',t)W_1(x'',t)\rangle \right\}, \nonumber
\end{align}
where we neglect correlations between the light and the
reservoir, of the type $\langle\psi_{in}^{\dagger}W_i\rangle,\cdots$. We use the properties
\begin{align}
\langle {\cal F}_1^{\dagger}(x',t'){\cal
  F}_2^{\dagger}(x'',t'')\rangle&=\langle {\cal F}_2(x',t'){\cal
  F}_1(x'',t'')\rangle=0,\nonumber \\
\langle
{\cal F}_1^{\dagger}(x',t'){\cal F}_1(x'',t'')\rangle&=\Gamma_a\bar{n}_a^{(1)}\delta(t'-t'')\delta(x'-x''),
\nonumber \\
\langle
{\cal F}_2(x',t'){\cal F}_2^{\dagger}(x'',t'')\rangle&=\Gamma_a(\bar{n}_a^{(2)}+1)\delta(t'-t'')\delta(x'-x''), \nonumber
\end{align}
where $\bar{n}_a^{(i)}$ is the average number of thermal phonons in the reservoir at
frequency $\Omega_i$. The expectation values are
\begin{align}
\langle W_1^{\dagger}(x',t)W_1(x'',t)\rangle&=\bar{n}_a^{(1)}\delta(x'-x'')\left(1-e^{-\Gamma_at}\right),\nonumber \\
\langle
W_2(x',t)W_2^{\dagger}(x'',t)\rangle&=(\bar{n}_a^{(2)}+1)\delta(x'-x'')\left(1-e^{-\Gamma_at}\right),\nonumber \\
\langle W_2(x',t)W_1(x'',t)\rangle&=\langle W_1^{\dagger}(x',t)W_2^{\dagger}(x'',t)\rangle=0, \nonumber
\end{align}
which lead to
\begin{align}
\langle\psi^{\dagger}(x,t)\psi(x,t)\rangle=\langle\psi_\mathrm{in}^{\dagger}(x-v_gt)\psi_\mathrm{in}(x-v_gt)\rangle
e^{-2Gx}+\frac{f_a^2}{2Gv_g^2}\left\{|{\cal E}_{1}|^2\bar{n}_a^{(1)}+|{\cal E}_{2}|^2(\bar{n}_a^{(2)}+1)\right\}\left(1-e^{-2Gx}\right)\left(1-e^{-\Gamma_at}\right). \nonumber
\end{align}
We interest in the limit of $GL\ll1$, where $L$ is the waveguide length. Then
the photon density, which is the number of
photons per unit length, is
\begin{align}
\langle\psi^{\dagger}(L,t)\psi(L,t)\rangle\approx\langle\psi_\mathrm{in}^{\dagger}(L-v_gt)\psi_\mathrm{in}(L-v_gt)\rangle+\frac{f_a^2L}{v_g^2}\left\{|{\cal
  E}_{1}|^2\bar{n}_a^{(1)}+|{\cal E}_{2}|^2(\bar{n}_a^{(2)}+1)\right\}\left(1-e^{-\Gamma_at}\right). \nonumber
\end{align}

\end{document}